\documentclass[aps,amsmath,amssymb,reprint,groupedaddress]{revtex4-1}

\usepackage{graphicx,times,amsmath,amsthm} 
\usepackage{appendix}
\usepackage[]{epsfig}
\usepackage{color}
\usepackage{xcolor}
\usepackage{bm}
\input{Qcircuit}

\newtheorem{theorem}{Theorem}[section]

\newtheorem{definition}[theorem]{Definition}

\begin{document}

\title{ Evolutionary Quantum Logic Synthesis of Boolean Reversible Logic Circuits Embedded in Ternary Quantum Space using Heuristics}


\author{Martin Lukac}
\email{lukacm@ecei.tohoku.ac.jp}
\altaffiliation{Department of Information Sciences, Tohoku University, Sendai, Japan}
\author{Marek Perkowski}
 \email{mperkows@ee.pdx.edu}
\altaffiliation{Department of Electrical Engineering, Portland State University, Portland, OR, USA}
\author{Michitaka Kameyama}
 \email{kameyama@ecei.tohoku.ac.jp }
\altaffiliation{Department of Information Sciences, Tohoku University, Sendai, Japan}

\begin{abstract}
It has been experimentally proven that realizing universal quantum gates using higher-radices logic is practically and technologically possible. We developed a Parallel Genetic Algorithm that synthesizes Boolean reversible circuits realized with a variety of quantum gates on qudits with various radices. In order to allow synthesizing circuits of medium sizes in the higher radix quantum space we performed the experiments using a GPU accelerated Genetic Algorithm. Using the accelerated GA we compare heuristic improvements to the mutation process  based on cost minimization, on the adaptive cost of the primitives and improvements due to Baldwinian vs. Lamarckian GA. We also describe various fitness function formulations that allowed for various realizations of well known universal Boolean reversible or quantum-probabilistic circuits.
\end{abstract}

\maketitle


\section{Introduction}

\noindent The nature of quantum particles and quantum environment allows quantum elementary particles to have in general more than two states~\cite{dirac:84, feynman:96}. Depending on the quantity that is being measured, the particles can have two, three, high natural number or even an infinity of quantum states. Depending on the technology used, these particles can be atoms, molecules, photons or potentially even electrons could make suitable candidates. However so far, most of the Quantum Logic Synthesis methods implicitly assume that the underlying mechanism of quantum computing is quantum-binary~\cite{khlopotine:02, lukac:02, miller:04, shende:05a, dueck:03, hung:04}.

\noindent This assumption is correct for classical technology where the radix 2 of a logic is the easiest to implement; for instance in CMOS logic, all values of a given logic are constrained to a finite range of allowed voltage. Thus building a multi-valued CMOS logic requires to place the desired number of logic values on the same allowed voltage range. This ultimately makes such implementations impractical as well as technologically infeasible. 

\noindent In Quantum mechanics, a similar limitation can be found. For instance, it is well known that a quantum system can have degenerate eigenstates; such states have the same value of a given measured quantity. However, in nature, many of the degrees of freedom of elementary particles will have more than two distinct and non degenerate states. The measured quantity can be energy, spin, position of a particle and so on. An example of an elementary particle with various degrees of freedom is a photon. It is currently used for various experiments and is also one of the most promising areas of quantum experimental computing~\cite{pachos:02, nielsen:03, xiaoqin:03, obrien:04, browne:05, fiurasek:06, kok:07, lanyon:08, fiurasek:08, gong:08}. A photon has a large number of degrees of freedom; these are polarization, transverse spatial-mode, arrival-time, photon number, and frequency~\cite{lanyon:08}. Other elementary particles offer large number of degrees of freedom with many states as well. For instance, trapped ions have electronic and vibrational energy levels with higher dimensions.

\noindent  Thus, assuming that at least k out of n possible states in a degree of freedom in an elementary particle can be attained at a reasonable cost of energy, it can be highly advantageous to use these higher-radix states for the synthesis of quantum logic circuits. Using this approach and using photons, a initial implementation of the Toffoli-sign (TS) gate using only three two-qubit quantum gates was experimentally demonstrated~\cite{lanyon:08}. This work was later extended into realizing generalized Toffoli gates using various methods and approaches such as in \cite{ionicioiu:09}. The TS gate approach is similar to the one in~\cite{nielsen:00} (page 183), but this time only with three CNOT gates and four single qubit rotation gates. Moreover, this approach scales better for larger number of control bits~\cite{lanyon:08} (linear versus quadratic in the number of two-qubit quantum gates).  

\noindent Evolutionary Algorithms allow to explore a very large problem space without knowledge of global structure of the problem space. In other words, an evolutionary algorithm can be used when the knowledge about the problem landscape is almost unknown and can be specified as a constraint satisfaction problem~\cite{eiben:03}. The required knowledge is in the local fitness function, and as such it means that one needs to know how a solution to the optimization problem is computed. 

\noindent The Evolutionary Quantum Logic Synthesis (EQLS) has been explored from various points of view in the last decade. On one hand the Genetic Programming has been widely used to synthesize quantum circuits and logic functions~\cite{rubinstein:01, leier:04, massey:04, massey:05, spector:04, stadelhofer:08}. On the other hand GA based methods have been applied to various quantum circuits as well~\cite{yabuki:00, lukac:02,lukac:03,lukac:09phd,lukac:08}. In general the focus of these approaches is either on a particular function (Reversible, quantum-permutative) or a general approach is used to see how well the evolutionary approach deals with this difficult problem. Several problems in EQLS have already been studied and analyzed, some of them are complexity of the quantum search space, high dimensionality of the quantum space, large number of quantum gates, etc. From these previous studies, it can be concluded that Evolutionary methods are well suited for research aimed to discover novel principles and novel quantum gate realizations of moderate size~\cite{lukac:09phd}. Following this reasoning, in this paper the focus is on the discovery and a deeper understanding of dynamics of the EQLS under various experimental conditions in the qudit framework.

\noindent In this paper we extend our initial work~\cite{lukac:10wcci} of EQLS of Boolean reversible and quantum-probabilistic circuits on qudits of higher dimensions ($d = 3$) using structural restrictions and heuristics. We provide a more detailed analysis of evolutionary mechanisms in the heterogeneous quantum logic synthesis and we compare results of the EQLS process with various improvements. The particular improvements to the Genetic Algorithm that we study are the following:
\begin{itemize}
\item Comparison of two types of fitness function 
\item Comparison between a Baldwinian and Lamarckian GA
\item Comparison between normal and adaptively weighted primitive sets
\end{itemize}

\noindent The experiments are performed over a selected set of both reversible and quantum gates and discussion of the results is provided.

\noindent This paper is structured as follows. Section~\ref{sec:1} describes the quantum circuits and the underlying quantum computing properties. Section~\ref{sec:2} describes the GA used in this work. Section~\ref{sec:3} presents and discusses the experimental results and Section~\ref{sec:4} concludes this paper.


\section{Quantum Gates and Quantum Circuits}
\label{sec:1}

\subsection{Boolean Quantum Gates}

The quantum gates used in this paper are some of the well known Boolean Quantum gates. The used quantum gates can be separated into permutative and non-permutative quantum gates. 

\subsubsection{Permutative Quantum Gates}
The permutative gates are also known as logically reversible quantum gates. Permutative gates in this work are the NOT (X) single qubit gate, CNOT and SWAP the two qubit permutative gates and Toffoli (also known as CCNOT), Fredkin (also known as CSWAP), Majority, Miller, Full Adder as the multi-qubit permutative quantum gates. The respective representation of the single and two qubit permutative quantum gates can be found in the Appendix~\ref{sec:app1}. Toffoli gate can be simply defined by eq.~\ref{eq:toffoli}
\begin{equation}
\begin{split}
CCNOT &= \ket{0}\bra{0}\otimes I^{\otimes^2} + \ket{1}\bra{1} \otimes CNOT\\
&= \ket{00}\otimes I+\ket{01}\otimes I+\ket{10}\otimes I+\ket{11}\otimes X 
\end{split}
\label{eq:toffoli}
\end{equation}

Fredkin gate is given by eq.~\ref{eq:fredkin}
\begin{equation}
CSWAP = \ket{0}\bra{0}\otimes I^{\otimes^2} + \ket{1}\bra{1} \otimes SWAP
\label{eq:fredkin}
\end{equation}

Majority is defined by the logic equation 
\begin{equation}
F = ab + bc + ac
\label{eq:majority}
\end{equation}
 and there are several implementations that do satisfy this logical requirement. Finally the Miller gate is given by eq.~\ref{eq:miller}
\begin{equation}
Miller = \begin{pmatrix} 1&0&0&0&0&0&0&0\\0&0&0&0&0&0&1&0\\0&0&1&0&0&0&0&0\\0&0&0&1&0&0&0&0\\0&0&0&0&1&0&0&0\\0&0&0&0&0&1&0&0\\0&1&0&0&0&0&0&0\\0&0&0&0&0&0&0&1\end{pmatrix}
\label{eq:miller}
\end{equation}

Finally, the full-adder is defined by two outputs as three inputs and its logic definition is shown in eq.~\ref{eq:fulladd}.
\begin{equation}
\begin{split}
S &= a\oplus b\oplus c\\
C &= majority(a,b,c)
\end{split}
\label{eq:fulladd}
\end{equation}

\subsubsection{Non-Permutative Quantum Gates}
The non permutative quantum gates used are the Z, Hadamard and their are shown in eq.~\ref{eq:z} and eq.~\ref{eq:hadamrd} respectively.
\begin{equation}
Z = (\ket{0}\bra{0} - \ket{1}\bra{1})*I
\label{eq:z}
\end{equation}

\begin{equation}
H = \frac{1}{\sqrt{2}}[\ket{0}\bra{0} +\ket{0}\bra{1} + \ket{1}\bra{0} - \ket{1}\bra{1}]
\label{eq:hadamrd}
\end{equation}
For the multi-qubit non-permutative gates the Controlled-Z (CZ) or the Controlled-H (CH) gates are used as well. The CH gate is given by
\begin{equation}
CH = \ket{0}\bra{0}\otimes I +\ket{1}\bra{1}\otimes H
\label{eq:chadamrd}
\end{equation}
and the CZ gate is given by
\begin{equation}
CZ = \ket{0}\bra{0}\otimes I +\ket{1}\bra{1}\otimes Z
\label{eq:cz}
\end{equation}

\subsection{Ternary Quantum Gates}

\noindent Circuits considered in this paper and in the original work of Lanyon~\cite{lanyon:08} contain the gate denoted by $[0-2]$. This is a multi-valued permutation quantum gate. Its matrix is shown in equation~\ref{eq:02permutation}(a). This quantum gates permutes the values of a single qutrit so that when the qutrit state is $|0\rangle$ it will result in $|2\rangle$ and if it is $|2\rangle$ it will become $|0\rangle$. Equation \ref{eq:02permutation}(b) shows a similar permutation gate $[1-2]$. This gate has similar functionality as the $[0-2]$ gate but swaps the states $|1\rangle$ and $|2\rangle$. Finally eq.~\ref{eq:02permutation}(c) shows the $[0-1]$ permutative gate. These three gates are the only truly multi-valued quantum gates used in this work.
\begin{equation}
\begin{tabular}{lc}
$[0-2]=\ket{0}\bra{2} + \ket{1}\bra{1} + \ket{2}\bra{0}$&(a)\\ 
$[1-2]=\ket{0}\bra{0} + \ket{1}\bra{2} + \ket{2}\bra{1}$&(b)\\
$[0-1]=\ket{0}\bra{1} + \ket{1}\bra{0} + \ket{2}\bra{2}$&(c)
\end{tabular}
\label{eq:02permutation}
\end{equation}

\subsection{Cost of Quantum Gates}

\noindent On of the criterion that the designed circuits are evaluated on is their cost. In this paper we are assuming a very simple quantum cost model. This model has been used for the standard quantum Boolean logic synthesis in various previous works~\cite{maslov:03, lukac:03, miller:04} and is only applied to single qubit and two qubit quantum gates. all other gates are built from these primitives. Thus, in this paper all two qubit/qutrit quantum gates have a cost of 1 and single qubit gates are ignored.

\subsection{Boolean Quantum Circuits Embedded in Qutrit Quantum Space}


\noindent 
Embedded Boolean quantum gates are different from multi-valued gates by the fact that as introduced in~\cite{lanyon:08}, these gates behave like regular Boolean gates for Boolean values of the input qutrit but for all other values of the qutrits they behave as identity operators.


\noindent For instance the Controlled-Z gate embedded in ternary quantum space is shown in eq.~\ref{eq:cz3}. 
\begin{equation}
CZ^3 = (\ket{0}\bra{0}+\ket{2}\bra{2}) \otimes I_3+ \ket{1}\bra{1} \otimes Z^3
\label{eq:cz3}
\end{equation}

\noindent Observe that the $CZ^3$ gate changes the phase of the target qutrit only when the input state is $\ket{11}$; this is the same function as a classical Boolean Controlled-Z gate.  On all other values of both the target and the control qutrit it must operate as an Identity operator~\cite{lanyon:08}.



An example of a non-permutative quantum gate embedded in a ternary quantum space is shown in eq.~\ref{eq:hadamard3}. 

\begin{equation}
H^3=\begin{pmatrix}\frac{1}{\sqrt 2}& \frac{1}{\sqrt 2}& 0\\\frac{1}{\sqrt 2}& -\frac{1}{\sqrt 2}& 0\\ 0&0&1\end{pmatrix}
\label{eq:hadamard3}
\end{equation}

In this article, the gates embedded in ternary quantum space are denoted by a an exponent of the embedding space; for instance a SWAP gate embedded in ternary space is denoted by SWAP$^3$ (eq.~\ref{eq:swap_3}).
\begin{equation}
SWAP^3 = (\ket{0}\bra{0}+\ket{2}\bra{2}) \otimes I_3 + \ket{1}\bra{1} \otimes X
\label{eq:swap_3}
\end{equation}

\subsubsection{Toffoli Gate}
\label{sec:toffoli}

\begin{figure}[h]
\centering
\includegraphics[width=0.5\textwidth]{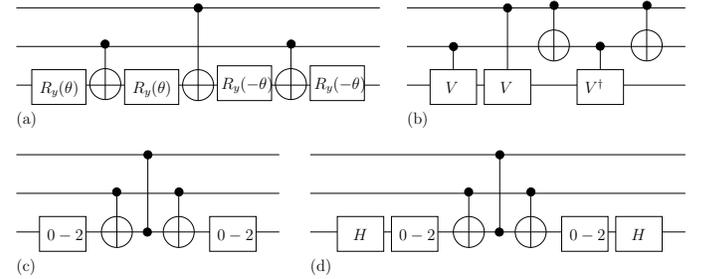}
\caption{\label{fig:toffolis} (a) Toffoli gate with a relative phase difference~\cite{nielsen:00}, (b) Toffoli gate realized using the CV/CV$^\dagger$ gates, (c) The Toffoli-Sign gate using multi-valued gates, (d) Toffoli gate constructed from Toffoli-Sign gate.}
\end{figure}

\noindent As shown in papers~\cite{ralph:07, lanyon:08} that are the prime motivation for this work, a Toffoli-Sign (TS) gate is such a gate that it changes the sign for only one particular input minterm (Figure~\ref{fig:toffolis}(c)). The TS gate become a logical Toffoli gate when two additional Hadamard gates are placed before and after the TS gate on the target qubit (Figure~\ref{fig:toffolis}(d)). 
\begin{table}[h]
\centering
\caption{\label{tab:toffolisign} The K-map of the Toffoli-sign gate: (a) using the $[0-2]$ gate, (b) using the $[1-2]$.}
\begin{tabular}{ccc}
\begin{tabular}{|c||c|c|}
\hline
$\qquad$c&0&1\\
ab& &\\
\hline
\hline
00&000&001\\
\hline
01&010&011\\
\hline
11&110&111\\
\hline
10&100&-101\\
\hline
\end{tabular}&&
\begin{tabular}{|c||c|c|}
\hline
$\qquad$c&0&1\\
ab& &\\
\hline
\hline
00&000&001\\
\hline
01&010&011\\
\hline
11&110&-111\\
\hline
10&100&101\\
\hline
\end{tabular}\\
(a)&&(b)
\end{tabular}
\end{table}
In a similar manner a Toffoli gate different from the logical Toffoli function by a relative phase from a normal Toffoli gate can be constructed using the approach shown in Figure~\ref{fig:toffolis}(a). In the Figure~\ref{fig:toffolis}(a) the rotations around the \em y  \em axis of the Bloch sphere use $\theta = \frac{\pi}{4} +n*\frac{\pi}{2}$~\cite{nielsen:00}. Finally, another and the most common realization of Toffoli gate - using the \em Controlled-V/V$^\dagger$\em is shown in Figure~\ref{fig:toffolis}(b).

\noindent Note that the Toffoli gate from Figure~\ref{fig:toffolis}(d) swaps the binary states $[101,100]$. So it works differently from normal Toffoli gate that swap states $[110,111]$. The Toffoli that swaps $[110,111]$ states is realized using the same principle but by replacing the $[0-2]$ gate by $[1-2]$ ternary gates (eq.~\ref{eq:02permutation}). In such case the realized gates swaps effectively the binary states $[110,111]$ (Table~\ref{tab:toffolisign}). The two-qubit gate in the middle of the circuit Figure~\ref{fig:toffolis}(c) and (d) is a Controlled-Z (CZ) gate.

\noindent Finally observe that the Toffoli gate in Figure~\ref{fig:toffolis}(d) works because the embedded Boolean quantum gates do not mix the ($\ket{0}$ ,$\ket{1}$) and the $\ket{2}$ quantum subspaces. In fact, it first separates the Boolean space of the target qutrit into two separate quantum spaces ($\ket{0}$, $\ket{1}$) and ($\ket{2}$) and then apply a single phase to one of the states in the ($\ket{0}$, $\ket{1}$) quantum space.

\begin{figure}[h]
\centering
\includegraphics[width=0.5\textwidth]{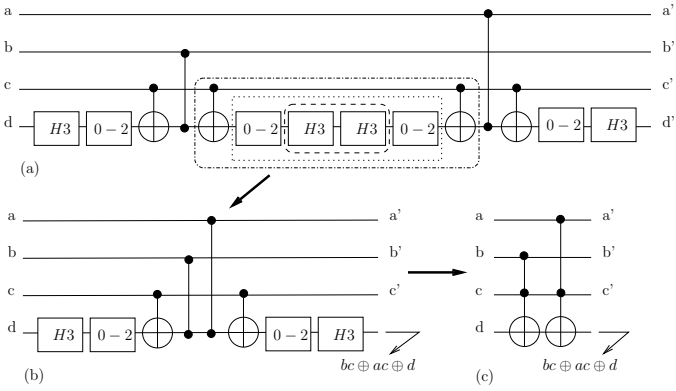}
\caption{\label{fig:boolefunction} (a) The realization of a Boolean function using Controlled-Z, controlled-NOT, Hadamard and $[0-2]$ multi-valued quantum gates, (b) the minimization of the number of quantum gates, (c) the Boolean function realized with two Toffoli gates.}
\end{figure}

\noindent The Toffoli gates introduced here all work on generally different principles from one another: we say that such circuits are realized using \em heterogeneous quantum gates \em. With respect to classical Quantum Logic Synthesis we define the \em heterogeneous quantum logic synthesis \em as: 
\begin{definition}[Heterogeneous Quantum Logic Synthesis]
A process of designing Quantum circuits using quantum gates of a various radices, acting on both the phase space and on the observable space. 
\end{definition}

\noindent Observe that unlike in classical Logic Synthesis, the heterogeneous quantum logic synthesis is a quite natural approach; no different technology and not distinctively different computational control protocol are required. 

\begin{figure}[h]
\centering
\includegraphics[width=0.5\textwidth]{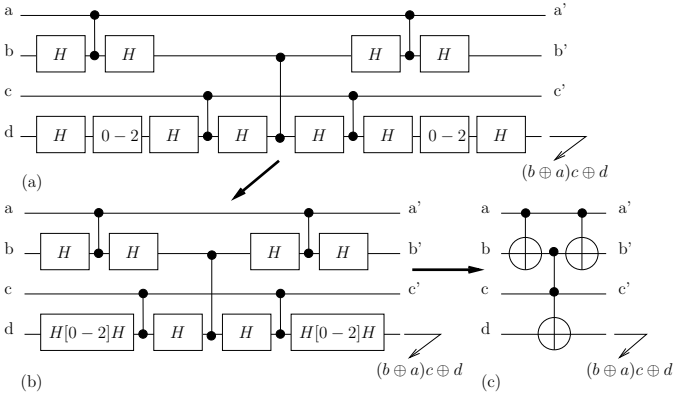}
\caption{\label{fig:anotherboolefunction} (a)The realization of a Boolean function using only Controlled-Phase, Hadamard quantum-binary gates and the $[0-2]$ multi-valued quantum gate, (b) the minimized circuit, (c) the same Boolean function realized using only Toffoli and Controlled-NOT gates.}
\end{figure}

\noindent As an illustration, the Figures~\ref{fig:boolefunction} and \ref{fig:anotherboolefunction} show the implementation of the $F_1=bc\oplus ac\oplus d$ and of the $F_2=(b\oplus a)\cdot c$ functions respectively. Observe that both circuits use the heterogeneous logic synthesis. 

\noindent The circuit from Figure~\ref{fig:boolefunction} shows how adjacent self-inverse gates are eliminated - the same process as in standard quantum Boolean logic. Similarly Figure~\ref{fig:anotherboolefunction} illustrates the gate concatenation based on the same principles as in qubit logic.

\noindent The $CZ^3$ gate can also be built as a mixture of qubit (control) and qutrit (target) and in such case its unitary matrix will be reduced from $3^3 * 3^3$ to $6 * 6$ rows-by-columns.

\noindent The boolean quantum logic synthesis uses two spaces: the phase space and the observable space. The difference between these two spaces can be seen in comparing the bases states. For instance, on one hand the common quantum bases states are $\ket{0}$ and $\ket{1}$. On the other hand more complex basis states are given by $\frac{\ket{0}+\ket{1}}{\sqrt{2}}$ and $\frac{\ket{0}-\ket{1}}{\sqrt{2}}$. Observe that applying the NOT operator will change state $\ket{0}$ to $\ket{1}$ and vice versa but the state $\frac{\ket{0}+\ket{1}}{\sqrt{2}}$ will change into $\frac{\ket{0}-\ket{1}}{\sqrt{2}}$ by using the Z quantum gate. This means that in the former set of basis states, the operator changes the value while in the later set of basis states the same operator changes only the phase.

\section{Genetic Algorithm for Quantum Logic Synthesis}
\label{sec:2}
\noindent The GA that was used for the experiments is an extension of our previous work \cite{lukac:03, lukac:08}. It designs Boolean quantum circuits embedded in multi-valued quantum space by only evaluating the Boolean input-output either by measuring for the two observables bases $\ket{0}$ and $\ket{1}$ or by comparing the coefficients of the circuit matrix with the target circuit matrix. For clarity we discuss only the most relevant features of the evolutionary search. For a complete description of the used GA the reader should consult~\cite{lukac:03, lukac:05, lukac:08, lukac:09phd}.

\subsection{The evolutionary process of computation}
\noindent The process of using GA for EQLS is described in the pseudo code below:
\begin{verbatim}
0: Initialize the GA and the primitive set
1: Generate Initial population of circuit
3: Evaluate the circuits in the population
2: Until(solution is found 
     or maximum generation is reached)
3:    select and replicate the circuits 
4:    mutate the circuits
5:    evaluate the circuits
6:    replace old population by the new one
7:    Goto 2
\end{verbatim}


\subsection{Primitive Set}

\noindent The GA builds circuits from a set of gates specified by the user called the \em primitive set\em. The size of this primitive set determines in a large extent whether yes or no the target circuit is found~\cite{lukac:09phd}. This is natural as without any restrictions, each additional gate that can be placed in a quantum circuit of a given size increases the possible combinations of the gates exponentially. The primitive set used in this work contains the following quantum gates: 
\begin{enumerate}
\item Controlled-NOT (CNOT$^3$), Controlled-Phase (CZ$^3$), $[0-2]$, $[1-2]$, Controlled-Hadamard (CH$^3$), Controlled-$[0-2]$, Controlled-$[0-1]$, Hadamard$^3$ and Wire.  
\end{enumerate}

\noindent The GA uses all primitive gates as they are; the control qubit will be always on the wire above the target qubit if the primitive gate is given as such. The only operation that the GA does on the gates is the expansion so that a CNOT gate can be placed on qubits that are not direct neighbors. 

\noindent The reason to restrict the primitive sets not to include the X single qubit gate is the fact that the $[0-1]$ quantum gate is the X gate embedded in the ternary quantum space as well as because of the fact that $H*Z*H = X$. Finally, as it was shown in our previous experimental work in EQLS the smaller primitive gate set is the higher is the probability that the desired circuit can be found. 

\begin{figure}
\centering
\includegraphics[width=0.30\textwidth]{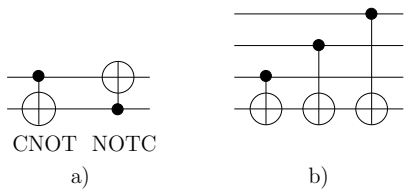}
\caption{Example of CNOT gates placement and uniqueness: a) The CNOT and the NOTC quantum gates, b)Example of CNOT gates requiring a unique encoding in the genotype string of a four qubit circuit}
\label{fig:encoding}
\end{figure}

\subsection{Circuits encoding}

\noindent The quantum circuits are presented as strings, each gate is uniquely defined by a triplet of characters. The reason for representing each gate by a triplet of character is purely algorithmic; when designing circuits on many qubits, the number of possible gates grows up quickly and thus a single character encoding does not allow to represent all required gates. As will be seen during the description of the Lamarckian Genetic Algorithm (Section~\ref{sec:darwlamarck}) it is required to store a large amount of gates that each require a unique character representation and a single character encoding (8 bits) allows to maximally represent 256 different quantum gates. For example a CNOT gate defined on two adjacent qubits and the same gate defined on two non adjacent qubits will have a different representation in the genotype string. Thus even with a finite set of input primitives, the number of unique representations grows with the width of the circuit. Figure~\ref{fig:encoding}(b) shows the different CNOT gates requiring different encoding in a circuit with four qubits. 

The eq.~\ref{eq:string} shows an example of a circuit represented as a string. For clarity, spaces have been added between each triplet of characters. 
\begin{equation}
C = p\; \underline{!!!}\; \underline{\#!!}\; pp\; \underline{\$!!}\; pp\; \underline{"!!}\; \underline{!!!}\; \underline{!!!}\; pp
\label{eq:string}
\end{equation}

\noindent A quantum circuit can be separated into a set of serially connected blocks, each acting on the same number of qutrits; each of the block acts on the exactly same number of qutrits as the whole circuit acts on. In eq.~\ref{eq:string} such blocks are delimited by a 'p' character from each side. Each such block is a sub-circuit acting on the same number of qutrits as the whole circuit acts on. The evolutionary algorithm operates on the circuit by either recombining the strings (exchanging the serial blocks) or by mutating elements within these blocks. According to the user-provided parameters, the mutation operator can be structure preserving (when one gate is replaced by another gate with the same number of inputs and outputs), not structure-preserving (a gate is changed into an arbitrary other gate) or a whole block can be erased and replaced by a new randomly generated block. In this experiment the mutation operator is not structure preserving but the probability of mutation is quite low. We use the Stochastic Universal Sampling replication mechanism~\cite{baker:87} and the two-point crossover. 


\subsection{Structural Restrictions}

In non-heterogeneous evolutionary quantum logic synthesis several parameters affect directly the success of the synthesis \cite{lukac:05a, lukac:08, lukac:09phd}. In particular a large number of input primitive gates reduces the chance to find the circuit~\cite{lukac:09phd}. This is particularly true when designing circuits with many quantum gates; circuits that require long sequences of gates - building blocks - are difficult to be maintained through the evolutionary process~\cite{eiben:03} and such sequences are generally lost when the crossover evolutionary operator is applied. This problem was experimentally shown as having a major impact on the EQLS performance~\cite{lukac:05, lukac:08, lukac:09phd}.

In the heterogeneous quantum logic synthesis there is additionally a third space: the radix space. To evolve quantum circuits in a heterogeneous quantum space additional quantum gates are required. This means that the initial primitive set is enlarged by not only Boolean quantum gates embedded in a higher radix quantum space but by also possible multi-valued quantum gates.  

However, as was introduced in Section~\ref{sec:toffoli} the heterogeneous quantum space allowed the design of a novel very low cost Toffoli gate and thus synthesizing circuits in this quantum space could allow the GA to find novel solutions to some of the well known quantum circuits.

In order to deal with the larger input primitive set required for the heterogeneous quantum synthesis a set of heuristics are introduced in this paper.

\noindent The first heuristic improvement introduced in this paper are the so-called \em structural restrictions\em. A structural restriction is a condition imposed on a quantum gate that allows this gate to be located or not to be located on a particular qudit of the quantum circuit. In the experimentations described later in this paper, some of the gates are allowed to be placed only on a specific wire in the quantum circuit. These restrictions are introduced on the observations made from the analysis of the circuits realized in the heterogeneous quantum logic synthesis~\cite{lanyon:08, lukac:10wcci}.


\noindent The Toffoli gate realized in this manner requires the multi-valued gates to be located only on the bottom (target) qubit. Using such knowledge about the Toffoli gate designed by Lanyon et al.~\cite{lanyon:08}, we can verify how fast the Toffoli gate will be designed by the GA without and with the structural restrictions. In this case the restrictions are on all single qubit gates but not on the Wire (Identity) single qubit gate. Thus for the primitive set, the single-qubit gates Hadamard and $[0-2]$ can be forced to be placed only on the bottom qudit (output qudit) of the quantum circuit. 

\begin{figure}[bth]
\centering
\includegraphics[width=0.3\textwidth]{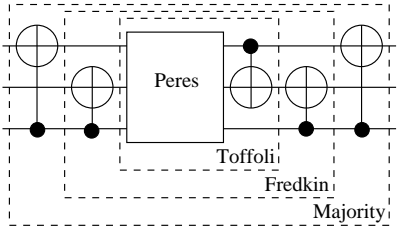}
\caption{\label{fig:ptf} The PTF principle}
\end{figure}

\begin{figure}[h]
\centering
\includegraphics[width=0.5\textwidth]{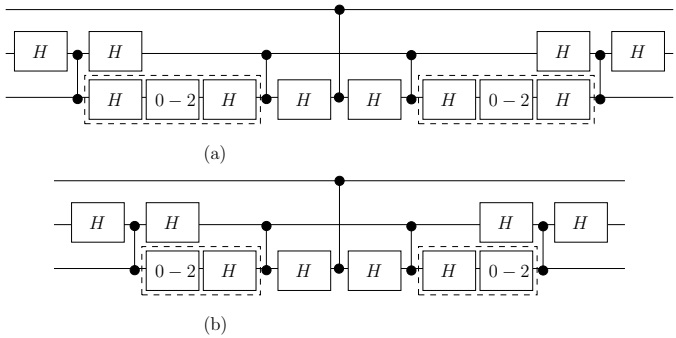}
\caption{\label{fig:fredkin}(a) Fredkin gate realized using the PTF principle, (b) CNOTC gate realized using Toffoli-sign and two NOTC gates.}
\end{figure}

\noindent To realize for instance a Fredkin gate one can use the Peres-Toffoli-Feynman (PTF) principle (Figure~\ref{fig:ptf}). This means that using two CNOT gates around a Toffoli gate will create a Fredkin gate. In the heterogeneous logic synthesis a possible solution is the multi-valued swap gate called SWAP$^3$. This gate is shown in eq.~\ref{eq:mv_swap}.
\begin{equation}
S_3 = \bordermatrix{
&00&01&02&10&11&12&20&21&22\cr
00&1&0&0&0&0&0&0&0&0\cr
01&0&0&0&1&0&0&0&0&0\cr
02&0&0&0&0&0&0&1&0&0\cr
10&0&1&0&0&0&0&0&0&0\cr
11&0&0&0&0&\underline{1}&0&0&0&0\cr
12&0&0&0&0&0&0&0&\underline{1}&0\cr
20&0&0&\underline{1}&0&0&0&0&0&0\cr
21&0&0&0&0&0&1&0&0&0\cr
22&0&0&0&0&0&0&0&0&\underline{1}
}
\label{eq:mv_swap}
\end{equation}
\noindent Using the gate from eq.~\ref{eq:mv_swap} as a part of a controlled gate the Fredkin realization could potentially be cheaper than in binary quantum. Figure~\ref{fig:fredkin} shows one possible realization of the Fredkin gate; Figure~\ref{fig:fredkin}(a) shows the Fredkin gate designed using the PTF principle (with Controlled-Phase gates, Hadamard and the ternary $[0-2]$) 
and Figure~\ref{fig:fredkin}(b) shows the Fredkin gate after gate minimization.

\begin{figure}[h]
\centering
\includegraphics[width=0.2\textwidth]{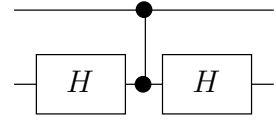}
\caption{\label{fig:feynman}CNOT gate realized using a CZ and two Hadamard gates.}
\end{figure}

\noindent It is also an open question if the multi-valued swap gate S$_3$ (eq.~\ref{eq:mv_swap}) can be constructed and what is its cost. So far using the approach studied in this article, it seems that because the CNOT gate (Figure~\ref{fig:feynman}) can be realized using only the Controlled-Z and Hadamard gate, the PTF principle seems to be the most natural. The search for novel realization of Feynman, Toffoli or Fredkin gates is one of the challenges of this paper. 

\begin{figure}[h]
\centering
\includegraphics[width=0.3\textwidth]{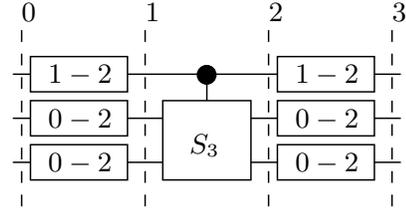}
\caption{\label{fig:fredkinm} Fredkin realized using a controlled multi-valued S$_3$ gate.}
\end{figure}

\noindent The multi-valued swap gate S$_3$ can also be realized using the approach proposed by Muthukrishan and Stroud~\cite{muthukrishnan:00, khan:07}. Thus a Fredkin gate can be also realized as shown in Figure~\ref{fig:fredkinm}. 

\noindent This can be shown by analyzing the input-output mapping of the Fredkin gate realization. This Fredkin gate inputs binary inputs and generates binary outputs. The Controlled-S$_3$ gate is activated only when the top qubit is in state 2; this is obtained by applying a rotation $[1-2]$ to the control qubit. The bottom two target qubits are first projected into the $\ket{1}$, $\ket{2}$ quantum space and then are swapped. To obtain back the binary values the bottom qubits are transformed using again the single $[0-2]$ rotation gates. 

Observe that this realization uses only multi-valued quantum gates; the Controlled-S$_3$ gate is only active when the control qubit is in state $\ket{2}$. Because the input data is restricted to binary values, the usage of the Controlled-S$_3$ gate is restricted to four input values only (Underlined coefficients in the matrix from eq.~\ref{eq:mv_swap}). This means that the gate Controlled-S$_3$ gate is designed to provide correct input-output mapping only when the control qubit is in the $\ket{2}$ state and the target qubits are in the quantum space spanned by $\ket{1}$ and$\ket{2}$. When the control qubit is in states $\ket{0}$ or $\ket{1}$ the Controlled-S$_3$ gate is not active and thus acts as an Identity gate but only on the target qubits being in the $\ket{1}$ and$\ket{2}$ quantum states.


\noindent  Thus the design of such gate is much simplified. The only required input-output mappings in the S$_3$ are now much easier to design as they represent only $\frac{2}{3}$ of the diagonal elements of the whole gate; this is essentially the same principle as simply embedding boolean quantum space to a ternary quantum space. As the input qubits will never be in the $\ket{1}$ for the control qubit and $\ket{0}$ for the target qubits the designed quantum gate is not required to be identity for these values from the logical point of view. Thus the gate S$_3$ is required to be only functionally corresponding to a swap gate only for the qubit values that will be available at its inputs or outputs.In this case these states are $\ket{21}$, $\ket{12}$, $\ket{11}$ and $\ket{22}$.

Naturally, this representation of Fredkin gate is equivalent to a Controlled-SWAP gate realized in higher-radix logic as long as it swaps two distinct values exclusively. Thus the circuit from Figure~\ref{fig:fredkinm} could be simply realized by a Controlled-Swap gate in a higher-radix logic assuming it is defined for the $\{100,101,110,111\}$ as a SWAP gate and for $\{000,001,010,011\}$ as the Identity gate. Such gate is a realization of quantum multiplexer in logic of higher dimension (equation~\ref{eq:CS12}). The don't care sign "-" represents the fact that the control qubit will never be in the state $\ket{1}$ and thus from the logic point of view it is not important what sub-matrix is defined for control qubit in state $\ket{1}$.

\begin{equation}
CS_{12} = \begin{pmatrix} \begin{pmatrix}I\end{pmatrix} &0 &0\\0& \begin{pmatrix}-\end{pmatrix}& 0\\0& 0&\begin{pmatrix}SW_{12}\end{pmatrix} \end{pmatrix}
\label{eq:CS12}
\end{equation}


Observe that all the gates discussed above and below are unitary because of the following synthesis restrictions imposed:
\begin{itemize}
\item All gates are designed using primitive unitary single and two qubit/qutrit quantum gates.
\item All primitive quantum gates embedded are also unitary as required by the initial condition imposed in~\cite{lanyon:08}
\item Gates not strictly obeying the unitary requirement are such gates that are unitary on all the available input-outputs mapping. This naturally means that the unused part of the quantum space (as the sub-matrix denoted by "-" in Figure~\ref{fig:fredkinm}) must also be unitary.
\end{itemize}

\noindent In a similar manner it can be expected that the majority gate can be realized using the heterogeneous EQLS. The majority gate has the advantage that it requires only one output; the reversible generalization of majority gate was proposed by~\cite{miller:02} and is called the Miller gate. The Miller gate can be directly obtained using the PTF principle by surrounding a quantum realized Fredkin gate by two more CNOT gates. However, in this paper we are interested in a novel implementation of the majority/Miller gate using the multi-valued quantum gates. The majority gate outputs a binary value "1" when the number of ones in the input is larger than the number of zeros. The output of the Majority gate is also given using eq.~\ref{eq:majority}.


\noindent The goal of designing the majority gate is to verify whether the structural restriction impact will be observable on a single-qubit output reversible function. In the case of the Fredkin gate, we are not aware about any direct solution using the quantum heterogeneous logic synthesis; the only known method of designing the Fredkin gate is the PTF principle that uses however only quantum Boolean gates. Thus the experiments will be conducted also over sets of restrictions in order to verify as many potential solutions as possible.

\begin{figure}[h]
\centering
\includegraphics[width=0.3\textwidth]{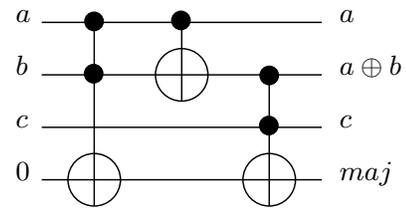}
\caption{\label{fig:majority} The majority gate realized using only MCT gates: $ab\oplus ac\oplus bc = ab\oplus (a\oplus b)c$.}
\end{figure}
\subsection{Baldwinian vs. Lamarckian Evolutionary Learning}
\label{sec:darwlamarck}

Before explaining the differences between Baldwinian and Lamarckian implementation of the EQLS we present some required basic concepts.

\subsubsection{Cost of Quantum Circuits}

\noindent One of the well known problems in Quantum Logic synthesis is how to find the quantum circuits with the minimal cost~\cite{lukac:03, maslov:03, lee:06}. The cost of an arbitrary quantum circuit is in general calculated from the number of quantum gates it contains. The procedure starts with assigning costs to the primitive gates and then for a given function the cost is the sum of costs of all gates. Let $g = \{g_0,\ldots,g_m\}$ be the set of all available quantum gates and $c=\{c_0,\ldots,c_f \}$ be the set of all possible costs of the quantum gates in $g$. Also let a function $r(g_p) = c_j$ be a mapping that takes a gate from $g$ as argument and returns the appropriate quantum cost from $c$. Then for any QC with k gates one can write $r(g^n) = c_j$ where $g_n \in g$ is the $n^{th}$ quantum gate from QC that corresponds to a unique gate in $g$ with the corresponding quantum cost $c_j$. Using this principle the cost of a QC is given by eq.~\ref{eq:quantumcost}.
\begin{equation}
C_{QC} = \sum_{j=1}^k r(g^j) = \sum_{j=1}^k c^j
\label{eq:quantumcost}
\end{equation}

\noindent with $c^j$ being the cost of the $j^{th}$ quantum gate in the circuit. On the lowest granularity level of quantum logic synthesis, the cost is given by  the number of single-qubit and two-qubit electro-magnetic (control) pulses. On this level the cost of well-known in terms of numbers of electromagnetic pulses have been computed in~\cite{lee:06}. Designing quantum circuits using control pulses is however quite difficult because even for small quantum gates such as Peres, Toffoli, Fredkin or Miller the number of control pulses reaches 12, 13, 19 and 24 respectively~\cite{lee:06}. Thus the design of a slightly larger quantum circuits will contain hundreds of the control pulses. From our previous experience with evolutionary quantum logic synthesis (EQLS), circuits with so many gates are difficult to synthesize because during the evolutionary process the building blocks are too large~\cite{lukac:05, lukac:08, lukac:09phd}. Thus, evolutionary synthesis should be not done on the level of control pulses.

\noindent In the most popular approach, the quantum cost is computed on the level of quantum logic gates; each logic gate such as Peres or Toffoli quantum gate is assigned a single cost and a gate is used as a macro in the synthesis process. Once a circuit is designed using these high level templates it is transformed to quantum counter parts - in general using the CNOT, CV and CV$^\dagger$ quantum gates and the final cost is computed after some local optimization on the level of control pulses or quantum primitives~\cite{lukac:03, maslov:03, miller:04, lukac:05}. However this approach is not guaranteed to produce minimal results on the level of quantum gates because it is not known if such transformations will always give the minimal quantum cost. For instance, using Multi-Controlled Toffoli (MCT) gates there are several high quality synthesizers~\cite{maslov:05, wille:09a}; however the synthesis using MCT gates does not guarantee an optimal (minimal) quantum cost.

\noindent Thus, with respect to the cost of quantum circuit the following dichotomy appears: the problem of Quantum Logic Synthesis is to find the most appropriate method that will avoid the high complexity space (control pulses)  and provide the best level of minimization.

\subsubsection{Correctness of Quantum circuit}
\label{sec:error}
\begin{definition}[Correctness of Quantum Circuit]
A quantum circuit is 100\% correct if all the logical elements of the circuit correspond to the desired target definition.
\end{definition}

For instance 80\% means that the given gate is correct in 80\% of the input-output mapping. The mappings used to evaluate the correctness of the quantum circuit can be either measured observables or the elements of the unitary matrix~\cite{lukac:09phd}

From logic point of view one can determine the correctness of a quantum circuit from three different perspectives:
\begin{itemize}
\item evaluating quantum circuit only on certain output qubits: this can lead to improved quantum cost but for the price that certain qubits remains in superposed quantum state~\cite{lukac:09phd}
\item evaluation of circuit correctness based only on a set of logic values, is equivalent to evaluation of the logical output after the measurement
\item evaluation of the circuit correctness based on matrix coefficient is the most exact method of evaluation but is not practically realizable as it requires to know the exact values of the matrix representing the unitary transform. 
\end{itemize}

In this work the correctness is evaluated on the logical values basis; the phase information of logic values is not taken into the account. 

\subsubsection{Input gate set reduction}

\noindent In EQLS the problem of synthesizing circuits with the smallest possible cost is difficult because when designing circuits on the level of control pulses  the primitive set is too large. This is equivalent to the statement that there are to many degrees of freedom in the selection of the primitive during the process of mutation. However when designing circuits with a small number of gates, the GA is outperformed by non-evolutionary approaches that use reversible logic synthesis design methods such as MMD algorithm~\cite{maslov:05} (bearing name after the authors initials) or Boolean Decision Diagram based~\cite{wille:09a}. However these approaches use well known and optimized techniques and thus are not suitable for the synthesis of arbitrary quantum circuits and do not allow discovering novel realizations of well known quantum or reversible functions.

\begin{figure}[h]
\centering
\includegraphics[width=0.3\textwidth]{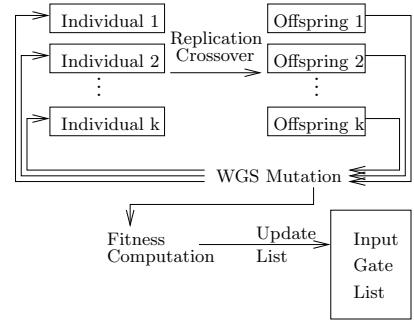}
\caption{\label{fig:adaptivegates} The process of ranking primitive gates in order to allow faster GA convergence}
\end{figure}

\noindent The approach proposed here is aimed to improve the ability of the GA to converge faster when designing quantum circuits with larger numbers of primitive gates. The process is illustrated in Figure~\ref{fig:adaptivegates}. 
\noindent The idea behind this adaptive process is the following: take the gates from the primitive set and assign to each of them a \em usage  \em and a \em fitness \em values. \em Usage \em represents how often the given gate is used while the \em fitness \em represents the average value of combined fitness of all circuits this gate has been used in. Thus each time a gate is used both the \em usage \em and the \em fitness \em value are updated and the gate is re-ranked within the pool of the primitive gates.  This dynamically growing and ranked primitive gate set is referred to as the \em extended primitive gate set \em (EIGS). Thus the gates that are created as replacement during the mutation process are now being selected using the \em Weighted Gate Set (WGS) mutation \em operator.

\noindent To implement the WGS mutation operator a variable \em gfitness \em is calculated and holds the sum of all gate-fitness values. The WGS mutation operator randomly selects a quantum gate from the circuit to be changed and then selects a new gate that matches both the structural restrictions as well as $r*gfitness$ value where $0\leq r\leq 1$ is a randomly generated number.
This approach allows to at least partially bias the selection of the replacement gates with three selection possibilities: trying new gates, trying good gates or trying any available gates from the primitive set.

\subsubsection{Baldwinian and Lamarckian Circuit evaluation}

\begin{figure}[h]
\centering
\includegraphics[width=0.3\textwidth]{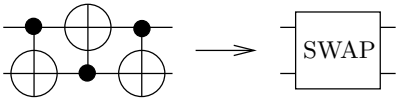}
\caption{\label{fig:swap1} Merging gates on the same quantum wires creates new larger gates that have been or can be optimized once for all on the control pulse level~\cite{lee:06}.}
\end{figure}

\noindent The Lamarckian EQLS is based on the well known minimization of the quantum gates. The general rule is given by:
\begin{definition}
Two adjacent quantum gates can be combined into a single quantum gate if they are acting on the same qubits.
\label{def:def2}
\end{definition}
\noindent For instance the well known realization of the two-qubit SWAP gate is based on three adjacent CNOT gates (Figure~\ref{fig:swap1}). This SWAP gate can be used as an optimized unit on the level of control pulses. Using this approach many quantum circuits can be minimized. Moreover and more important for EQLS is the fact that the merging of gates on one hand reduces the total number of quantum gates in the quantum circuit and on the other hand increases the number of quantum gates in the primitive gate set. This process of gate merging is also shown in Figures~\ref{fig:boolefunction} and ~\ref{fig:anotherboolefunction}.

\noindent This means that when starting form an initial set of quantum gates soon a very large set of quantum gates is available for the mutation operator. Thus using the EIGS and the WGS mutation operator it is possible to partially control which gates are reused and which can be omitted. The combination of EIGS and of the WGS mutation operator is the base of the Lamarckian and Baldwinian algorithms.

\noindent The difference between the two approaches (Baldwinian and Lamarckian GA) is in the fact that once a circuit is constructed, it is minimized using the above rules and in the case of the Lamarckian EQLS the new genotype replaces the old one. In the case of the Baldwinian EQLS only the cost and the fitness calculation is done using the optimized circuit representation but the genotype remains unchanged. Thus, in the Lamarckian GA, the actual genotype of the individual is rewritten by the minimized string and new gates are actively created and inserted into the primitive gate set (Row 5 in the Table~\ref{tab:darwvslamar}).

\noindent One of the possible problems when using the proposed Lamarckian EQLS is that if it is not bounded by some hard imposed limits the EIGS would explode in the computer memory. This problem would make the GA practically impossible to run as it would run out of physical memory within few initial generations. The other problem is that, even if we impose a hard limit to the extended set of gates, it is hard to replace the least fit gates: the gates that have been least used and have the lowest gfitness value in this run. This means, that one could imagine an approach where with a limited set of extended quantum gates, the least fit of them are replaced by the newly generated. The problem with this approach is that it would require that all the removed gates are also removed from the population. 
\noindent The alternative approach to this was to impose a hard limit on the number of quantum gates and reduce the number of iterations during the GA run and average the extended set of gates over multiple runs that ended in the same local or global maximum. This allowed us to evaluate the overall convergence of the extended gate set. In the presented experiments the EIGS size was limited to 50 quantum gates.

\begin{table}[h]
\centering
\caption{\label{tab:darwvslamar} Comparisons of differences between Baldwinian and Lamarckian EQLS}
\begin{tabular}{|c|c|c|c|}
\hline
&Lamarckian & Baldwinian& Classical\\
\hline
\hline
Mutation& WGS & Standard&Standard\\
\hline
Crossover & Standard & Standard&Standard\\
\hline
Input gate set& Ranked, dynamic& Initial&Initial\\
\hline
Genotype& Replaced & Preserved&Preserved\\
\hline
Phenotype& Optimized& Optimized&Standard\\
\hline
\end{tabular}
\end{table}

\noindent Table~\ref{tab:darwvslamar} sums up the main differences between the Baldwinian, the Lamarckian and a classical EQLS. The classical GA does not use any optimization for calculating the cost of the quantum circuits. The Standard key word refers to evolutionary operators that operates on well-known principles in the GA. The Ranked and dynamic primitive gate set means that the EQLS uses the EIGS: the EIGS  is growing with novel gates being created and inserted into it, and is dynamically updated using the fitness and usage of the gates in real time. Initial primitive gate set means that the primitive gate set selected by the user does not change during the whole evolutionary process. Observe that for both Baldwinian and Lamarckian GAs the computed fitness values are calculated from the optimized genotype.

\subsection{GPU acceleration}

\noindent As already described, in this work, all qubit operators are embedded in ternary operators in such manner that for one value of the qudit they act like the Identity.  
Similarly the circuits evolved are in principle ternary but only subsets of desired input and output values are used for the actual computation and evaluation processes.

\noindent Thus the overhead of computation can be quite large. This computational overhead is the main motivation to use a dedicated GPU computational processor for the evaluation of quantum circuits in the EQLS. The GA uses the GPU based matrix computation in order to accelerate the overall computation by taking advantage of the parallelism in the CUDA~\cite{cuda} libraries. The GPU acceleration is used to compute the matrix of the circuit. In general the GPU comes with multiple cores and each core is organized into a two-dimensional (can be virtually organized into a three-dimensional) grid of process threads. In order to implement an optimal matrix multiplication several conditions must be satisfied. Most important is that the grid of the computational processes should be allocated optimally. It means that the matrix sizes should be multiples of the number of the parallel computational processes. 

\noindent The second constraint of similar importance is that in order to execute GPU computations the data must be transported from the main memory to the GPU memory using the data bus. As this bus is used also for other devices on the computer, it is possible to saturate the bandwidth and thus actually slow down the computation. 

\noindent The GA used in this work is configured to compute ternary quantum matrices. Such matrices cannot be computed in the most optimal approach in a GPU device that is in general designed to have $2^k$ computational threads. However even in such case this computation is more efficient than on the CPU as it allows to unload the CPU to perform other tasks in parallel. Also to save the data bus bandwidth the computation of the circuit matrix representation is done in two steps:

\begin{enumerate}
\item Compute the matrix of the circuit by sending all required parallel blocks (matrices) one by one to the GPU
\item Return the computed resulting matrix.
\end{enumerate}

\section{Experiments and Results Discussion}
\label{sec:3}

\noindent 
The reversible quantum gates that we attempted to synthesize are the Toffoli gate(Toffoli-Sign), the Fredkin gate, the majority gate, the Miller gate, the SWAP$_3$ gate and the full adder. All gates can be synthesized using the GA when using only the Boolean quantum gates and thus present good benchmarks to evaluate the performance of our Genetic Algorithm. The reason for selecting this set of gates is that they represent a very basic standard of well known and used universal reversible gates. 

\noindent The experiments are evaluated for up to 10000 generations; this number was previously experimentally determined as to be sufficient to observe if a given gate can be found in a statistically significant manner~\cite{lukac:09phd}. All resulting data are represented as average over 50 runs. The default fitness function used is the $f_1$ fitness function from eq.~\ref{eq:fit1} unless specified explicitly. 

\noindent Two fitness functions have been used for the experimentation. The $f_0$ function is a simple fitness function given by eq.~\ref{eq:fit1}:
\begin{equation}
f_0 = \frac{1}{1+error}
\label{eq:fit1}
\end{equation}
\noindent where $error$ being given by the square difference between the desired and the target circuit output. A more elaborate fitness function that takes into account the cost of the quantum circuit is shown in eq.~\ref{eq:fit2}:
\begin{equation}
f_1 = \alpha\frac{1}{1+error}+\beta\frac{1}{cost}
\label{eq:fit2}
\end{equation}
\noindent with $\alpha$ and $\beta$ are constants obeying unity given by $\alpha+\beta = 1$ and $cost$ being the cost of the circuit in the range $[1,\inf]$. The values of the coefficients $\alpha$ and $\beta$ are experimentally determined and are not the subject of study here. However as can be expected if the influence of the cost is too high the synthesized circuits will be of a smaller cost but will rarely represent the correct function. 

\noindent The first set of experiments is aimed to show the difference between the usage of the restrictions. For this certain gates from the primitive set have been restricted to only certain wires. These gates are the $[0-2]$, $[1-2]$ and the $H^3$ and in the restrictions allowed them to be placed only on a single qubit at a time. The results of both runs are reported and compared in Table~\ref{tab:results}. Our algorithm was able to find solutions to most of the benchmark functions in the allocated time and space. Each row in Table~\ref{tab:results} shows the benchmark function name in the first column, and the number of generations if a solution was found for GA run with the structural restrictions (column 2) and without restrictions (column 3). The column entitled "Corr.", corresponds to average correctness of the target gate evaluated using the error described in Section~\ref{sec:error}. For instance the Toffoli gate was generated in 500 generation in average when using the structural restrictions and with the average correctness of 95\% (in 5\% of the cases the correct gate was not found and the algorithm stopped because reached the maximum allowed generations).

\begin{table}[h]
\centering
\caption{\label{tab:results} Overview of the results of the synthesis problem}
\begin{tabular}{|c|c|c|c|}
\hline
Gate&\multicolumn{2}{|c|}{Number of Generations}& \\
Name& w. Restrictions& wo. Resctrictions& Corr.\\
\hline
Toffoli& 500& 2500& 100\%\\
\hline
Majority&500 &2000& 97\%\\
\hline
Fredkin&1500 & -&95\%\\
\hline
SWAP$_3$&500 &1000&98\%\\
\hline
Miller& 10000& -&90\%\\
\hline
\end{tabular}
\end{table}

\noindent Observe that the structural restrictions allow to design the given circuit considerably faster in the case where the solution was found. In the cases where the algorithm did not find any solution, the number in the third column indicates the percentage of correctness of the given gate. 


\noindent The application of structural restrictions proved to be useful as it allows to directly reduce the time required to find solutions to our benchmarks. Naturally as can be expected, one needs prior knowledge about at least the principles of the desired quantum gates in order to be able to produce a set of restrictions that speed up the evolution. If a wrong set of restrictions is applied then the evolution can both take longer or will not find the target gate at all. 

\noindent For instance. using the principle proposed in~\cite{lanyon:08} one could restrict the usage of single qudit quantum rotation gates (depending on the number of control qudits~\cite{lanyon:08}) to only the wires where the actual output function is created. This however works only for the MCT gates. All MCT gates have only a single functional output while all other inputs are just passing through and are unchanged on the output. However, the same restrictions will not be successful on a permutative non MCT gate as for instance is the Miller gate; for this gate the restrictions should be less strict as the circuit realizes majority function on each of the output qubits. In such case the gates should not be using structural restrictions. 

\begin{table}[h]
\centering
\caption{\label{tab:lamdarnor}Comparison of results when using the Lamarckian GA with the primitive gate set ranking side by side with the Baldwinian GA and the standard not improved GA.}
\begin{tabular}{|c|c|c|c|}
Gate& Classical& Lamarckian and WGS& Baldwinian\\
Name& (Gen./Corr.)& (Gen/Corr.)& (Gen/Corr.)\\
\hline
Toffoli& 1500/95\%& 900/100\%& 1600/95\%\\
Majority& 1000/95\%& 750/100\%& 1200/95\%\\
Fredkin& -&1700/100\% & 2000/96\%\\
Full Adder& 120/97\%&250/100\% &180/98\% \\
\hline
\end{tabular}
\end{table}

\noindent The next set of benchmarks illustrates the convergence changes as a consequence of using the Lamarckian and the Baldwinian learning. Table~\ref{tab:lamdarnor} shows the comparison of the three main types of GA tested: the standard (column 2) the Lamarckian (column 3)  and the Baldwinian (column 4).  Each of the three columns (2, 3 and 4) shows the average number of generations until the target gate was obtained and the average correctness of the obtained result. Observe that using the Lamarckian learning significantly boosted the evolutionary process.

\noindent It is particularly interesting to notice that in fact the Baldwinian GA does not really improve the performance of the GA. The reason for this is that because the actual size of the circuits is not modified. Thus during mutation and recombination, same restrictions applies to the Baldwinian GA as to the standard one. The fitness and the cost of the minimized circuit has no effect on the genetic operators and thus as the main constraint is the size of the circuit with the right combination of quantum gates, the Baldwinian GA is not more successful.

\noindent Table~\ref{tab:lamvdar} shows the comparison between the Lamarckian and the Baldwinian evolution. The Lamarckian GA is evaluated with and without using adaptive weighted primitive gate set. Observe that as expected the performance of such Lamarckian GA without the use of the EIGS is much worse than any other GA configuration used.
\begin{table}[h]
\centering
\caption{\label{tab:lamvdar}Comparison of results when using the Lamarckian GA with and without the primitive gate set ranking side by side with the Baldwinian GA}
\begin{tabular}{|c|c|c|c|}
Gate& Lamarckian& Lamarckian and WGS& Baldwinian\\
Name& (Gen./Corr.)& (Gen./Corr.)& (Gen./Corr.)\\
\hline
Toffoli& 2100/80\%& 900/100\%& 1600/95\%\\
Majority& 2000/70\%& 750/100\%& 1200/98\%\\
Fredkin& 2500/85\%& 1700/100\%& 2000/95\%\\
Full Adder& 420/97\%& 250/100\%& 180/98\% \\
\hline
\end{tabular}
\end{table}

\noindent Moreover, the performance of the Lamarckian GA without the use of the EIGS is actually much worse than the performance of the standard non-improved GA. This can be seen when comparing the second column of Table~\ref{tab:lamvdar} with the first column of Table~\ref{tab:lamdarnor}. 

\noindent The above benchmarks all used cost related fitness functions from eq.~\ref{eq:fit1}. To observe how the fitness function affects the convergence, the following experiments are focused on the correctness of the synthesized quantum circuit (error) and its cost. The Table~\ref{tab:fitness} shows how the two main classes of fitness functions affect the convergence of the algorithm. The first column of the Table~\ref{tab:fitness} shows the name of the synthesized function. The second column in Table~\ref{tab:fitness} illustrates the performance of the GA using the fitness function given by the eq.~\ref{eq:fit1}. The third column in Table~\ref{tab:fitness} shows the same data but using the fitness function from eq.~\ref{eq:fit2}. 
\begin{table}[h]
\centering
\caption{\label{tab:fitness}Comparison of results when using the normal fitness function vs. using fitness function with cost as a component of the fitness value.}
\begin{tabular}{|c|c|c|}
Gate Name& Fitness $f_0$& Fitness $f_1$\\
& (Cost/Corr.)& (Cost/Corr.)\\
\hline
Toffoli& 8/95\%& 7/87\%\\
Majority& 10/95\%& 9/90\%\\
Fredkin& 15/40\%& 12/60\%\\
Full Adder & 10/97\%& 15/98\%\\
\hline
\end{tabular}
\end{table}

\noindent  Columns two and three in Table~\ref{tab:fitness} show the average cost and the average correctness of the target circuit.

For instance the Toffoli gate row shows that using the fitness function $f_0$ (second column) the average cost of the Toffoli gate was 8 and the average correctness of the synthesized Toffoli gate is 95\%. This means that on the 50 runs of the GA in 2 runs the obtained gate was not the exact match to the desired Toffoli gate.

\begin{table}[h]\centering
\caption{\label{tab:efficiency}Comparisons of the rate of success of the GA under the various settings with the average cost of solutions found. The numbers represent averaged percentage of the successful target gate as well as the cost averaged over 20 GA runs.}
\begin{tabular}{|c|c|c|c|}
Gate Name& Standard& Lamarckian and WGS& Baldwinian\\
&Succ/Cost&Succ/Cost&Succ/Cost\\
\hline
Toffoli& 98\%/20& 100\%/18& 100\%/22\\
Majority& 90\%/40& 100\%/25& 100\%/29\\
Fredkin& -&50\%/25 & 20\%/30\\
Full Adder& 90\%/35&100\%/28 &93\%/35 \\
\hline
\end{tabular}
\end{table}

\noindent The last set of experiments is intended to demonstrate the improvements given by the Lamarckian GA; the Table \ref{tab:efficiency} shows the percentage of the GA success for the tested gate benchmarks. Observe that the Lamarckian GA using the WGS mutation allows to synthesize the circuits with a higher rate of success as well as with a lower cost. The three data columns in Table~\ref{tab:efficiency} represent the average success of the synthesis (Succ) of the target gate and the average cost (Cost).

\noindent As already partially described, when using Lamarckian learning, the problem is that the primitive gate set is actively modified. Because of this it is required to provide additional information about the newly created quantum gates otherwise it will be very difficult for the GA to find the correct circuit. This is because as shown in ~\cite{lukac:03,lukac:09phd} a large primitive set means that the appropriate set of gates to be used in the target circuit is harder to select as the probabilities of selection are distributed over a larger primitive set. Thus, in our case we add the gate-fitness that allows to implement a ranking of the quantum gates. This means, that as long as a list of available quantum gates can be maintained the GS mutation operator will allow to select a set of gates that have a higher probability to be present in the target final gate. 

\noindent It is also important to notice that the ranking of the gates depends on the initial population. Thus On one hand, a set of generated gates during one run will end up having a different ordering than when another run is performed. This is natural as which gates are preserved in the memory depends on the initial state of the population. On the other hand, for the same desired quantum gate to be synthesized, statistically the same gates will end up being included in the extended set of quantum gates. This is because, as the evolutionary process advances, the local maximum of fitness also partially determines what gates are used in the circuit. 

\begin{figure}[h]
\includegraphics[width=0.5\textwidth]{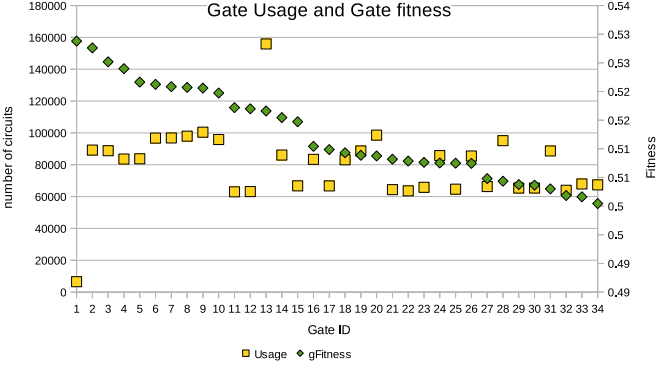}
\caption{\label{fig:gatesstats}The final result of the classification of the extended gate set. Usage represents the number of times a given gate has been used in a circuit, and \em gfitness \em is the average of the fitnesses of all circuits it has been part of.}
\end{figure}

\noindent Figure~\ref{fig:gatesstats} shows gates from a single run for the Fredkin quantum gate. Observe that the difference between the gates on the top of the list and on the bottom of the list is relatively small. The peak at gate ID 12 represents the identity gate.

\noindent Unfortunately it is not practically possible to compare gate lists from various runs because the gates that are placed into the EIGS can have similar string encoding in different evolutionary runs but do not necessarily have the same representing unitary matrix. Thus a comparative representation would be meaningless. However, the graph from Figure~\ref{fig:gatesstats} can be used to set up preferential parameters for further investigations.

\begin{figure}[h]
\centering
\includegraphics[width=0.3\textwidth]{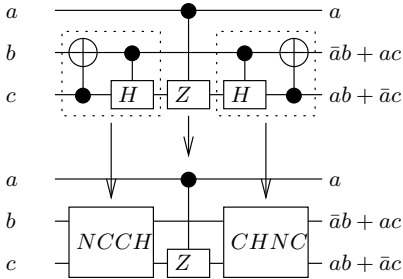}
\caption{\label{fig:fredkin_new} The discovered Fredkin quantum gate and its minimization. NCCH stands for NOT-Controlled Controlled-Hadamard and CHNC is Controlled-H NOT-Controlled.}
\end{figure}

\noindent The most important result from this search is the novel implementation of the practical Fredkin gate. The circuit is shown in Figure~\ref{fig:fredkin_new}. Despite the fact that this gate is realized using on qubit operators, it was discovered by the algorithm while searching for Fredkin gate with elementary gates used for the EQLS in the ternary quantum space. In other words, this gate does not use any properties of the qutrit quantum space to generate the qubit outputs.

\noindent The Fredkin gate from Figure~\ref{fig:fredkin_new} is in fact the cheapest realization so far known in the quantum logic synthesis. It is particular because it heavily relies on the phase kick-back from both the controlled-Z and from the controlled-Hadamard gate but does not use the principles of the heterogeneous quantum logic synthesis. Also, when minimized, using the concatenation rule from def.~\ref{def:def2}, this gate has the same cost as the Toffoli gate! This is due to the fact that the circuit given by the three middle gates is a Toffoli gate, but unlike in the methodology using the CV/CV$^\dagger$/CNOT gates, it does not require additional CNOT gate to return the control inputs to the initial values (Figure~\ref{fig:toffolis}b). Thus, it can be minimized to a smaller cost.

\begin{figure}[h]
\centering
\includegraphics[width=0.3\textwidth]{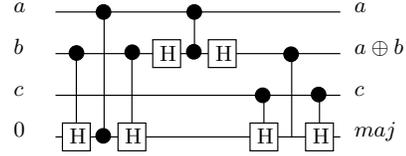}
\caption{\label{fig:majority_s} The majority quantum gate designed using the Controlled-H and the Controlled-Z gates. }
\end{figure}

\noindent Similarly the Majority gate was designed by the GA using the controlled-H and the controlled-Z gate. It is shown in figure~\ref{fig:majority_s}.

\begin{table}[htb]
\begin{tabular}{|c|c|c|c|}
Gate name& Best Boolean Cost& Best Calculated& Best Obtained\\
\hline
Toffoli&5&3&3\\
Fredkin&7&5&5\\
Miller&9&7&8\\
Majority&12&8&10\\
Full-adder&12&12&12\\
\hline
\end{tabular}
\label{tab:final_results}
\end{table}

\noindent Finally, to show the differences between the cost of quantum circuits designed using heterogeneous quantum logic synthesis the Table~\ref{tab:final_results} shows the best different coasts of the various quantum circuits. The second column of Table~\ref{tab:final_results} shows the best known quantum coast when quantum boolean gates only are used. The third quantum column shows the equivalent of the best quantum boolean realization when re-synthesized using the heterogeneous quantum synthesis. The fourth column shows the best cost obtained using the experimentation described in this paper. 

\section{Conclusion}
\label{sec:4}

\noindent First, in this paper we experimentally approached the heterogeneous QLS using an Evolutionary Algorithm with a set of structural restrictions and Lamarckian GA. Both of the proposed improvements to standard GA were beneficial on the benchmark quantum circuits that we were able to synthesize under the prescribed experimental conditions. As such, the Lamarckian GA and the structural restrictions are a quite successful improvement to the EQLS. In the described experiments only circuits of relatively small size have been explored. However even in these benchmarks the improvement is observable and  thus similar improvement of performance can be expected for large circuits composed from basic gates designed here. 

\noindent Second, we designed new set of quantum gates automatically: Toffoli (realization previously known), Fredkin (realization previously unknown), Adder (realization previously unknown) and Majority (realization previously unknown). The obtained results validate the proposed approach not only as a theoretical study but also shows that GA is very well situated for exploration of unknown problems in the QLS area.

\noindent The future work on this approach is to extended it to a truly hybrid heterogeneous QLS, where qubits, qudits and qutrits are truly mixed in a quantum circuit and its software representation. This has the advantage that the dimension of the circuit is significantly reduced; linearly in the size of the matrices. For instance a circuit with three qubits is represented by a $3^2\otimes3^2\otimes3^2 = 27^2$ complex coefficients. Assuming that two qudits in such circuit use only  $|0\rangle$ and $|1\rangle$ states, a circuit with two qubits and one qutrit has its matrix of size only $2^2\otimes2^2\otimes3^2 = 12^2$. Moreover, the mixing of qudits of different radices allows a faster selection of operators that can be applied and thus subsequently allows to create more and more sophisticated \em structural restrictions\em, the concept proposed in this article. The possible disadvantage is that the multiplication of operators for qudits of various radices creates matrices that might be not well suited to be computed quickly using the CUDA approach. The possible improvement of the GPU approach presented is to move all operators (matrices) to the GPU memory, and perform all matrix computations as well as error evaluations directly on the GPU. This means that during an evolutionary run there will be almost no information transmitted between the memory/CPU and GPU but short strings or scalar values.

\appendix
\section{Permutative Quantum Gates}
\label{sec:app1}

The NOT (also called X) gate is shown in eq.~\ref{eq:x}

\begin{equation}
X = \ket{0}\bra{1} + \bra{1}\ket{0} 
\label{eq:x}
\end{equation}

The CNOT gate is shown in eq.~\ref{eq:cnot}

\begin{equation}
X = \ket{0}\bra{0} \otimes I + \ket{1}\bra{1} \otimes X
\label{eq:cnot}
\end{equation}

\bibliographystyle{plain}
\small{
\bibliography{../main}
}

\end{document}